\documentstyle[emulateapj,apjfonts,psfig]{article}
\begin{document}

\title{\large \bf The Shape of Spectral Breaks in GRB Afterglows}

\author{ 
Jonathan Granot\altaffilmark{1}
and Re'em Sari\altaffilmark{2}}

\altaffiltext{1}{Racah Institute of Physics, Hebrew University,
Jerusalem 91904, Israel}

\altaffiltext{2}{Theoretical Astrophysics 130-33, California Institute
of Technology, Pasadena, CA 91125, USA}

\begin{abstract}
  Gamma-Ray Burst (GRB) afterglows are well described by synchrotron
  emission from relativistic blast waves expanding into an external
  medium. The blast wave is believed to amplify the magnetic field and
  accelerate the electrons into a power law distribution of energies
  promptly behind the shock. These electrons then cool both
  adiabatically and by emitting synchrotron and inverse Compton
  radiation. The resulting spectra is known to consist several power
  law segments, which smoothly join at certain break frequencies.
  Here, we give a complete description of all possible spectra under
  those assumptions, and find that there are 5 possible regimes,
  depending on the ordering of the break frequencies. The flux density
  is calculated by integrating over the contributions from all the
  shocked region, using the Blandford McKee solution. This allows us
  to calculate more accurate expressions for the value of these break
  frequencies, and describe the shape of the spectral breaks around
  them. This also provides the shape of breaks in the light curves
  caused by the passage of a break frequency through the observed
  band. These new, more exact, estimates are different from more
  simple calculations by up to a factor of $\sim 70$, and describe some new
  regimes which where previously ignored.
\end{abstract}


\section{Introduction}\label{sec:intro}

In recent years, several dozens of gamma-ray burst (GRB) afterglows
have been observed, and data is accumulating rapidly. The quality of
these observations is constantly improving. The study of afterglow
emission has helped shed light on many important aspects of the GRB
phenomenon. The spectrum during the afterglow phase is well described
by synchrotron emission from a relativistic blast wave, and consists
of several power law segments (PLSs), that join at several break
frequencies (e.g. Sari, Piran \& Narayan 1998). These break
frequencies are the self absorption frequency, $\nu_{a}$, below which
the optical depth to synchrotron self absorption is larger than unity,
$\nu_m$, the typical synchrotron frequency of the minimal electron in
the power-law, and $\nu_c$, the synchrotron frequency of an electron
whose cooling time equals the dynamical time of the system. Granot,
Piran \& Sari (2000) have then found that if $\nu_c<\nu_m$, the self
absorption frequency actually splits into two: $\nu_{ac}$ and
$\nu_{sa}$, where an optical depth of unity is produced by non-cooled
electrons and all electrons respectively. Different possible orderings
of these break frequencies result in five possible spectral regimes,
as shown in figure \ref{Fig1}.

The physical parameters of a burst may be deduced from fitting the
observed broad band spectrum to the theoretical spectrum. This has
been done by Wijers \& Galama (1999) for GRB 970508 by fitting a
broken power law theoretical spectrum.  A detailed description of the
shape of the spectrum allowed a more accurate determination of the
self absorption frequency $\nu_{sa}$ and the peak frequency $\nu_m$
(Granot, Piran \& Sari 1999b, hereafter GPS99b). A more accurate
theoretical calculation of the break frequencies leads to a more
accurate conversion from the observed spectrum to the burst
parameters. The combined effect was that the inferred value of the
density, for example, was different than that of Wijers and Galama by
two orders of magnitude. This illustrates the sensitivity of this
method to the shape of the theoretical spectrum around the break
points, and stresses the need for a more accurate determination of the
theoretical break frequencies, for all various spectral breaks.

So far only the shape of the spectrum around $\nu_m$ (Granot, Piran \&
Sari 1999a, hereafter GPS99a; Gruzinov \& Waxman 1999) and $\nu_a$
(GPS99b) was calculate in detail, and even that was only done for the
canonical case where $\nu_{sa}<\nu_m<\nu_c$ (see the upper panel of
figure \ref{Fig1}). This paper, is intended to extend these works for
{\it all} spectral breaks, and therefore provide a comprehensive, self
consistent calculation of the broad band spectrum. We provide analytic
formulas which approximate the shape of each of the spectral breaks
and their positions, in a form which is easy to use for afterglow
fitting. We also suggest a prescription for combining these breaks to
a single analytic broad band spectrum.

The physical model is outlined in \S \ref{PM}, while a more detailed
and formal description of the model and of the calculation of the
observed flux density is given in \S \ref{AA} (Appendix A).  Our main
results are presented in \S \ref{ANAP}. In \S \ref{BBS} we give
prescriptions for combining the shapes of the spectrum near the
different spectral breaks into a single analytic broad band spectrum.
We discuss our results in \S \ref{dis}.

\section{The Physical Model}
\label{PM}

An exact calculation of the spectrum requires the knowledge of (i) The
hydrodynamic quantities (bulk Lorentz factor and number density) (ii)
The magnetic field strength (iii) The electron energy
distribution. These should be given for any location behind the shock,
and at any time. Below we describe our approach to all three.

(i) The hydrodynamics is described by the Blandford McKee (1976, BM
hereafter) self similar solution. This solution describes a spherical
relativistic blast wave expanding into a cold medium, and assumes an
adiabatic flow, i.e. that radiation losses are small and do not effect
the hydrodynamics.  Radiative effects can be taken into account to
modify the hydrodynamic evolution as described by (Sari 1997, Cohen,
Piran \& Sari 1998) and to modify the structure of the cooling layer
behind the shock as described by Granot \& K\"onigl (2001).  If the
radiative losses are not too large, our formalism would give the
correct break frequencies and break shapes, provided that one uses the
time dependent energy, as discussed in the first two of these
references.  The BM solution we use is for an impulsive explosion in
an ambient density described by a power law with radius, $\rho_{\rm
  ext}(r)=A r^{-k}$. We consider two different values of $k$ which are
of particular physical interest: $k=0$, corresponding to an
interstellar medium (ISM), and $k=2$, corresponding to a massive star
progenitor, surrounded by its pre-explosion wind.  The assumption of a
spherical flow is also adequate for a jetted flow at sufficiently
early times, when the Lorentz factor of the flow is still larger than
the inverse opening angle of the jet. We therefore have the complete
hydrodynamic description in term of the total energy $E$, and the
external number density $n_{\rm ext}$ (or $A$ in the case of wind).
The hydrodynamic profile that is used, is given in equations
\ref{BM_e} through \ref{BM_n}.

(ii) We assume the magnetic field gets a fixed fraction, $\epsilon_B$,
of the internal energy everywhere behind the shock, as given by
equation \ref{B}. This would be the case if the shock amplified,
randomly oriented, magnetic field decreases due to adiabatic
expansion.  Different assumptions on the evolution and orientation of
the magnetic field were shown to have only a small effect on the
resulting spectrum (GPS99a,b).

(iii) The electrons are assumed to acquire a power law distribution of
energies, $N(\gamma)\propto\gamma^{-p}$ for
$\gamma\geq\gamma_{{\rm min},0}$, immediately behind the shock. Their total
energy immediately behind the shock is a fraction $\epsilon_e$ of the
internal energy. After being accelerated by the shock, the electrons
cool due to radiative losses and adiabatic cooling. The former can be
calculated from synchrotron theory and the latter from the density
profile given in by the BM solution. The resulting distribution 
is given in equation \ref{dist}.

Given the above, the observed flux density may be calculated as
described in \S \ref{AA} (Appendix A). The spectrum for the optically
thin breaks may be calculated using equation \ref{optically_thin},
while equation \ref{F_nu_tau>1} applies more generally.

\section{Results}
\label{ANAP}

\begin{figure*}
\centerline{\hbox{\psfig{figure=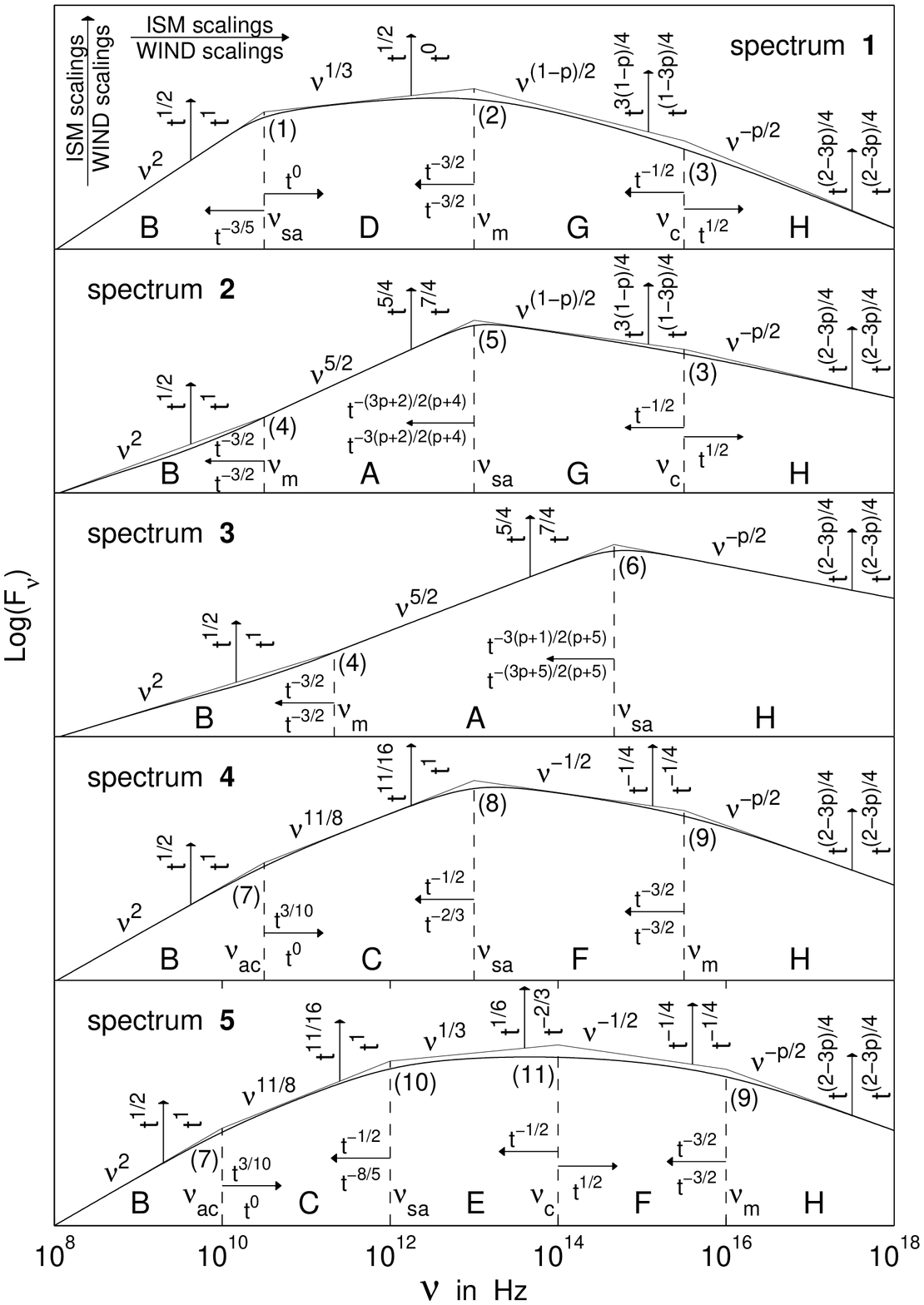,width=15cm}}}
\caption[]{\label{Fig1} 
  The different possible broad band synchrotron spectra from a
  relativistic blast wave, that accelerates the electrons to a power
  law distribution of energies. The thin solid line shows the
  asymptotic power law segments (PLSs), and their points of
  intersection, where the break frequencies, $\nu_b$, and the
  corresponding flux densities, $F_{\nu_b,{\rm ext}}$, are defined.
  The different PLSs are labeled A through H, while the different
  break frequencies are labeled 1 through 11. The temporal scalings of
  the PLSs and the break frequencies, for an ISM ($k=0$) or stellar
  wind ($k=2$) environment, are indicated by the arrows.  The thick
  solid line shows the spectrum we calculated in this paper, where the
  broad band spectrum is constructed according to the prescription
  suggested in \S \ref{BBS}. The different spectra are labeled 1 through
  5, from top to bottom. The relevant spectrum is determined by the
  ordering of the break frequencies. The top two panels (spectra 1 and
  2) correspond to slow cooling ($\nu_m<\nu_c$).  Spectrum 1 applies
  when $\nu_{sa}<\nu_m$, while spectrum 2 applies when
  $\nu_m<\nu_{sa}<\nu_c$.  The two bottom panels (spectra 4 and 5)
  correspond to fast cooling ($\nu_c<\nu_m$).  Spectrum 5 applies when
  $\nu_{sa}<\nu_c$, and spectrum 4 applies when
  $\nu_c<\nu_{sa}<\nu_m$. Spectrum 3 (middle panel) applies when
  $\nu_{sa}>\nu_m,\nu_c$, where in this case the relative ordering of
  $\nu_c$ and $\nu_m$ is unimportant (i.e. spectrum 3 may apply both
  to slow cooling or fast cooling).}
\end{figure*}

{\noindent \bf power law segments.}  All five possible different
spectra, which are shown in Figure \ref{Fig1}, consist of between
three and five different power law segments (PLSs). At each asymptotic
PLS (sufficiently far from the break frequencies) we have
$F_{\nu}\propto\nu^{\beta}t^{\alpha}$. Altogether there are eight
different PLSs, labeled A through H, from high to low values of
$\beta$. Note that there are two different PLSs with a slope of
$\beta=1/3$. Both are produced by the low energy tail of synchrotron
radiation, but in region $D$ it is the non-cooled electrons that are
responsible for the radiation while in region $E$ it is the cooled
electrons. Most PLSs appear in more than one of the five possible
spectra (see Figure \ref{Fig1}).  If one is only interested in the
spectrum far enough from the break frequencies, then the
normalizations of the different PLSs is all that is needed to
accurately describe the spectrum. This is given in Table \ref{PLS}.
The coefficients for PLSs A, G and H, slightly depend on $p$ in a non
analytic way.  These were calculated for $p=2.2,2.5,3$, and a linear
function (or a linear function multiplied by an exponent) was used to
describe the results. 

For PLS E, we find that the emission becomes dominated by the
contribution from small radii (i.e. early times when the radius of the
shock was small) for $k \geq 23/13\approx 1.769$. The electrons
responsible for the emission in this regime have suffered considerable
adiabatic cooling (as well as radiative losses). In this regime
($k>23/13$) PLS E splits into two different PLSs, whose spectral slope
$\beta$ depends on $k$. Furthermore, the effective size of the
afterglow image, at a given observed time, in this regime depends on
the observed frequency. However, since this new regime is somewhat out
of the main stream of this paper, and in order to avoid confusion, we
leave the detailed description of this new regime to a future work
(Granot \& Sari, in preparation). The normalization of PLS E, and the
expressions for the spectral breaks $b=10,11$ (that involve PLS E),
are therefor left out of Tables \ref{PLS} and \ref{T1}, respectively,
for $k=2$.

{\noindent \bf break frequencies.}  The different possible
combinations of the eight PLSs, result in 11 different break
frequencies, labeled $b=1,...,11$ (see Figure \ref{Fig1}). Again, the
same break frequency may appear in more than one spectrum. The values
of the break frequencies, $\nu_b$, and the corresponding extrapolated
flux densities, $F_{\nu_b,{\rm ext}}$, are defined at the point where
the asymptotic PLSs meet. These can be directly calculated from the
normalization of the PLSs that are given in Table \ref{PLS}, but for
completeness they are given explicitly in Table \ref{T1}. The fit for
the $p$ dependence was redone in this table (with either a linear fit,
an exponent or a combination of the two) using the accurate results
for $p=2.2,2.5,3$, therefore resulting in slight (a few percent)
inconsistencies with the previous table. 

{\noindent \bf shape of breaks.}
The Flux density near a spectral break, $\nu_b$, may be approximated
by
\begin{equation}\label{anapp}
F_{\nu}=F_{\nu_b,{\rm ext}}\times
\left[\left({\nu/\nu_b}\right)^{-s\beta_1}
+\left({\nu/\nu_b}\right)^{-s\beta_2}\right]^{-1/s}\ ,
\end{equation}
where $\beta_1$ and $\beta_2$ are the asymptotic spectral slopes below
and above the break, respectively, and $s$ is a parameter which
describes the sharpness of each break. The sign of $s$ is equal to
that of $\beta_1-\beta_2$ (i.e. positive if the spectral slope
decreases across the break), while $|s|$ represents the sharpness of
the break (the sharper the break, the larger $|s|$). The shape of most
spectral breaks (except for $b=1,8,10,11$), depends on the value of
$p$, and so does the corresponding value of $s=s(p)$, which is given
in Table 2. All quantities which depend on the value of $p$ were
calculated for $p=2.2,2.5,3$, and are given in a form which is as
exact as the functional parameterization permits at these values of
$p$, and interpolates (or extrapolates) for other values of $p$, and
should therefore be reasonably accurate for $1.5\lesssim p \lesssim
3.5$ (see discussion below equation \ref{electron_dis} for $p<2$).

The break $b=1$ has been investigated in detail by GPS99a, for $k=0$,
and they found that the physically motivated formula
\begin{equation}
\label{b1}
F_{\nu}=F_{\nu_{1},{\rm ext}}\left[1-\exp(-\tau_{1})\right] \tau_{1}^{-6/5}
\quad ,\quad \tau_{1}=(\nu/\nu_{1})^{-5/3} \ , 
\end{equation}
provides an even better description of $F_{\nu}$ near the break (with
an MRD of $2.63\%$, compared to $6.78\%$ with equation
\ref{anapp}). However, for $k=2$, equation \ref{anapp} provides a
better fit (with an MRD of $1.02\%$ compared to $25\%$ with equation
\ref{b1}), which shows that the previous success of equation \ref{b1}
was accidental.  

Equations \ref{anapp} and \ref{b1} both give a poor fit for
$b=4$. This is because the spectral slope across this break does not
change monotonically. We therefore provide an alternative formula for
this break, 
\begin{equation}
\label{b4}
F_{\nu}=F_{\nu_{4},{\rm ext}}\left[ \phi_4^2 \exp(-s\phi_4^{2/3})
+\phi_4^{5/2}\right]\quad , \quad \phi_4=\nu/\nu_4\quad \ ,
\end{equation}
where the values of $s$ for $b=4$ which appear in Table 2, are for
this equation, rather than for equation \ref{anapp}, as for the other
breaks.

\section{A Prescription For The Broad Band Spectra}
\label{BBS}

The values and the shape of the break frequencies, as given in the
previous section, are strictly valid only when the break frequencies
are far away from each other. Though, in principle, our
formalism is adequate to describe the general spectrum, for arbitrary
values of the break frequencies, such a description, would require a
new calculation for any ratio of the break frequencies, and is
therefore not practical.  Instead, we choose to give a heuristic
prescription, that uses the shapes from the previous section to
construct a broad band spectrum, which includes all the breaks, for an
arbitrary ratio of the break frequencies. Once again we stress that
this is not a rigorous derivation of such a spectrum, but simply an
analytic equation, that gives a smooth spectrum when the break
frequencies are close to each other, and approaches the rigorous shape
of each break, in the asymptotic situation where the break frequencies
are far apart. Such an equation is useful for fitting afterglow data.

One can readily construct such a formula for any one of the five
different possible spectra shown in Figure \ref{Fig1}. Let us label
these spectra 1 through 5, from top to bottom, and denote the
corresponding flux densities by $F_{\nu}^{(i)}$ where $i=1,...,5$. We
also label the flux density near the 11 spectral breaks by $F_{b}$,
where $b=1,...,11$. The fluxes, $F_{b}$, are given by equation
\ref{anapp} (for break $b=4$ equation \ref{b4} gives a more accurate
description). Now, let us define a quantity $\tilde{F}_{b}$, by
\begin{equation}\label{F_tilde}
\tilde{F}_{b}  = \left[1+\left({\nu/\nu_b}\right)^{s(\beta_1-\beta_2)}\right]^{-1/s}\ .
\end{equation}
The formulas for the rounded shape of the spectrum for the five
spectra which are shown in Figure \ref{Fig1}, from top to bottom, are
given by
\begin{eqnarray}\label{bbs}
F_{\nu}^{(1)} &=& F_{1} \tilde{F}_{2} \tilde{F}_{3} \ ,
\\
F_{\nu}^{(2)} &=& F_{4} \tilde{F}_{5} \tilde{F}_{3} \ ,
\\
F_{\nu}^{(3)} &=& F_{4} \tilde{F}_{6} \ ,
\\
F_{\nu}^{(4)} &=& F_{7} \tilde{F}_{8} \tilde{F}_{9} \ ,
\\
F_{\nu}^{(5)} &=& F_{7} \tilde{F}_{10} \tilde{F}_{11} \tilde{F}_{9} \ .
\end{eqnarray}
The first term, $F_b$, provides the normalization and the shape of the
spectrum near the lowest break frequency, while each consecutive term,
$\tilde{F}_b$, represents the next break frequency, from low to high
frequencies, and provides the shape of the spectrum near that break
frequency, and the appropriate change in the spectral slope across
the break.  The number of free parameters in each spectrum generally
equals the number of break frequencies plus two, since besides the
values of the break frequencies, one has to specify the value of $p$
and of the flux normalization. The bottom panel of figure \ref{Fig1}
is an exception, and $F_{\nu}^{(5)}$ has only 5 free parameters, since
there is a closure relation between the four break frequencies
(Granot, Piran \& Sari 2000):
\begin{equation}\label{closure_relation}
{\nu_{10}\over\nu_7}\left({\nu_{11}\over\nu_9}\right)^{4/5}=h(p)\sim
1 \quad ({\rm for}\ k=0)\ .
\end{equation}

Our prescription for constructing the broad band spectrum uses $F_{\nu_b,{\rm ext}}$ from only one of the break
  frequencies in each spectrum, and thus avoids the problem of the
  slight inconsistencies within Table \ref{T1} (that arise due to the
  independent fits for the $p$ dependence, e.g. \S \ref{ANAP}).

\section{Discussion}
\label{dis}

We have used the BM solution, to obtain more accurate expressions for the
flux density. Under the assumptions that the initial electron distribution is
a strict power-law with a low energy cutoff, and that the magnetic
field is amplified immediately behind the shock, we derived exact
expressions for the values of the break frequencies, as well as the
shape of the spectrum around each break. We have given a complete
general description of the broad band spectrum. As our analysis is
general, it also includes exotic spectra, that may only be relevant
in very early phases or for extreme parameters. Our main results are
summarized in Figure \ref{Fig1} and Tables 1 and 2.

In general, the spectrum of GRB afterglows evolves from fast to slow
cooling\footnote{This holds for $k<4$, which includes the cases
  relevant for the afterglow, $k=0,2$.}.  For example, for an ISM with
standard parameters, (e.g. $n_0\cong 1$,$E_{52}\cong 1$, $\epsilon_B
\cong 0.01$) the initial spectrum is 5, then $\nu_m$ crosses $\nu_c$
and the spectrum turns into spectrum 1, and finally, when $\nu_m$
crosses $\nu_{sa}$ the spectrum turns into spectrum 2.  The transition
times between the various spectra of Figure \ref{Fig1} can be worked
out by equating the various frequencies, as given in Table
\ref{transitions}. It follows that there are two types of evolution
depending on the parameters, as given in the chart below.
 
$$
{\rm ISM}=\cases{
5 \rightarrow 1 \rightarrow 2 & $n_0 E_{52}^{4/7} \epsilon_B^{9/7} < 18$ 
\cr 
4 \rightarrow 3 \rightarrow 2 & $n_0 E_{52}^{4/7} \epsilon_B^{9/7} > 18$  
}
$$
$$
{\rm WIND}=\cases{
4 \rightarrow 5 \rightarrow 1 \rightarrow 2 & $A_{*}\bar\epsilon_e\,^{-1}E_{52}^{-3/7}\epsilon_B^{2/7}>100$ \cr 
4 \rightarrow 3 \rightarrow 2 & $A_{*}\bar\epsilon_e\,^{-1}E_{52}^{-3/7}\epsilon_B^{2/7} $  < 100
} 
$$

The weakest link in our formalism, is the assumption of a sharp lower
cutoff in the electron distribution. This would affect breaks
$b=1,2,4,7,9$ (though for $b=1$, the shape of the break will not be
effected, while $\nu_1$ and $F_{\nu_1,{\rm ext}}$ may change).  
Nevertheless, our calculation provides the first self
consistent description of all these breaks.  The values and shape of
the rest of the breaks, depends only on the assumption of a power-law
distribution, well above the low energy cutoff, and on the electron
cooling. Our description of these breaks ($b=3,5,6,8,10,11$) is
therefore more robust. These breaks may still be somewhat affected by
the assumption of the magnetic field evolution. However, in previous
papers (GPS99a,b), we have shown that this dependence is relatively
week ($F_{\nu_2,{\rm ext}}$ and $\nu_2$ change by up to $\sim 50\%$,
while $F_{\nu_1,{\rm ext}}$ and $\nu_1$ change only by up to a few
percent, where in both cases the shape of the break does not change
considerably).

We give a complete description of all possible power-law segments
(PLSs), and provide exact expressions for the flux density away from the break
frequencies. These expression are useful when partial information for
the afterglow exists. In general a spectral slope and a flux at some
frequency are sufficient to give some constraint on the afterglow
parameters (in PLSs G and H, $p$ would also be needed). For example,
if only X-ray data exists, PLS H can be used to extract some
information on the underlying parameters, and if only radio data
exists, PLS B can be used, even if the self absorption frequency is not
observed, (i.e. is above the observed radio frequency).

Expressions for some of the break frequencies and corresponding flux
densities already exist in the literature (Waxman 1997; Sari, Piran \&
Narayan 1998; Wijers \& Galama 1999; GPS99a,b; Granot, Piran \& Sari
2000; Chevalier \& Li 2000; Panaitescu \& Kumar 2000). Most of these
works address the spectrum shown in the upper panel of Figure 1
(spectrum 1).  The values we obtain for the break frequencies and
corresponding flux densities are in some cases significantly different
than previous estimates (by up to a factor of $\sim 70$). For $k=0$
and $p=2.5$, our value for $\nu_2$ ($F_{\nu_2,{\rm ext}}$), which is
better known as $\nu_m$ ($F_{\nu_m}$), is a factor of 1.3 (4.2) larger
(smaller) than Sari, Piran \& Narayan (1998), a factor of 1.5 (3.1)
smaller (larger) than Wijers \& Galama (1999), a factor of 3.8 (for
$\nu_2$) smaller than Panaitescu \& Kumar, and a factor of 15 (8)
smaller (larger) than Waxman (1997). For $p=2.2$ our value for $\nu_2$
is a factor of 70 smaller than Waxman (1997)\footnote{This large
  difference is mainly due to the fact that Waxman used $\epsilon_e$
  instead of our $\bar\epsilon_e=\epsilon_e(p-2)/(p-1)$.}. Our values
for $\nu_2$ and $F_{\nu_2,{\rm ext}}$ are only slightly different (by
$-5.1\%$ and $+1.6\%$, respectively) than GPS99a, due to a small
approximation they made for the local emissivity.  Our value for
$\nu_3$ ($\nu_c$) is a factor of 2.6 larger than Sari, Piran \&
Narayan (1998), a factor of 6.4 larger than Wijers \& Galama (1999),
and a factor of 6.1 larger than Panaitescu \& Kumar. Our value for
$\nu_1$ ($\nu_{sa}$) is a factor of 1.9 larger than Waxman (1997), a
factor of 3.7 smaller than Wijers \& Galama (1999), a factor of 4.1
larger than Panaitescu \& Kumar, and a factor of 2.1 smaller than
GPS99b\footnote{The reason for this last difference is as follows.
  Equation 18 of GPS99b, which is essentially equation 6.52 of Rybicki
  \& Lightman (1979), misses the term associated with the
  discontinuity at the lower edge of the electron distribution (at
  $\gamma_{\rm min}$) when derived from equation 6.50 of Rybicki \&
  Lightman. This missing term caused an overestimation of the of the
  absorption coefficient by a factor of $f=3(p+2)/4$, and a
  corresponding overestimation of $\nu_1$ and $F_{\nu_1,{\rm ext}}$,
  by factors of $f^{3/5}$ and $f^{1/5}$, respectively. However, this
  missing term does not effect the shape of the break, which is given
  in equations \ref{anapp} or \ref{b1} (i.e. equation 24 of GPS99b).}.
For $k=2$ (and $p=2.5$), our values of $\nu_1$, $\nu_2$ and $\nu_3$
are smaller by factors of 2.5, 1.4 and 1.4, respectively, compared to
Chevalier \& Li (2000), while our value for $F_{\nu_2,{\rm ext}}$ is
larger by a factor of 3.2; Compared to Panaitescu \& Kumar, or value
for $\nu_1$ ($\nu_2$) is larger (smaller) by a factor of 4.9 (7.4),
while our value for $\nu_3$ is larger by only $16\%$. Our expressions
for the break frequencies and corresponding flux densities of spectra
4 and 5 (bottom two panels of Figure 1), for $k=0,2$, are different by
up to a factor of $3$ from those given in Granot, Piran \& Sari
(2000).

Our equations do not include the effects of inverse compton scattering
on the cooling of the electrons. This effect is known to be important,
when $\epsilon_B\ll \epsilon_e$ (Sari, Narayan \& Piran 1996,
Panaitescu \& Kumar 2000, Sari \& Esin 2001). Following the
prescription of Sari \& Esin, we can include the effects of inverse
compton by inserting appropriate powers of $(1+Y)$ into the values of
the break frequencies, or the PLSs (where $Y$ is the compton
y-parameter). PLSs C, E, F and H should be multiplied by $(1+Y)^{-3/8}$,
$(1+Y)^{2/3}$, $(1+Y)^{-1}$ and $(1+Y)^{-1}$, respectively.

Preliminary results from this work have already been used successfully
in fitting the data of several afterglows (e.g. Galama et al. 2000;
Harrison et al. 2001). In the latter case the first evidence for
inverse Compton emission was found. A special effort has been made to
present the results of our model in a way that is simple to implement,
and would provide the most accurate results to date for spherical
afterglows, or jetted afterglows within their quasi spherical phase
(before any significant lateral spreading).

\acknowledgements

JG thanks the Horowitz foundation for support. RS thanks the Sherman
Fairchild foundation for support. This research was partially
supported by a NASA ATP grant.

\section{Appendix A}
\label{AA}

The energy density $e$, number density $n$, magnetic field $B$, and
random Lorentz factor of the electrons $\gamma_e$, are measured in the
local rest frame of the fluid, in addition to all the primed quantities. The
remaining quantities are measured in the lab frame, i.e. the rest
frame of the ambient medium in which the flow is spherical. We use a
spherical coordinate system in this rest frame, where the z-axis
points at the observer. The time, $t$, measured in this rest frame is
called the coordinate time, and is to be distinguished from the time,
$t'$, measured in the local rest frame of the fluid, and from the
observer (or observed) time, $t_{\rm obs}$, at which the emitted
photons reach the observer.  The subscript `$0$' denotes the value of
a quantity just behind the shock.

The initial electron distribution, just behind the shock, is given by
\begin{equation}\label{electron_dis} 
N(\gamma_e)=K_0\gamma_e^{-p}\ \  {\rm for}\ \  
\gamma_e\geq\gamma_{{\rm min},0}=
{\bar\epsilon_e e_0\over n_0 m_e c^2}\ ,
\end{equation}
where $m_e$ is the electron rest mass and
$K_0=(p-1)n_0\gamma_{{\rm min},0}^{p-1}$. Note, that the above equation is
usually written using $\epsilon_e=\bar\epsilon_e(p-1)/(p-2)$, which
is the fraction of the internal energy given to the electrons. The
advantage of using $\bar \epsilon_e$, is that it makes most equations
somewhat simpler.  Furthermore, it will apply also for the case $p<2$,
as long as the minimal Lorentz factor is proportional to the shock
Lorentz factor. The magnetic field is assumed to hold a constant
fraction, $\epsilon_B$, of the internal energy, everywhere,
\begin{equation}\label{B} 
B^2 = 8\pi\epsilon_B e\ .
\end{equation}

The evolution of the Lorentz factor of each electron is described by
\begin{equation}\label{dgamma_dt} 
{d\gamma_e\over dt'}=-{\sigma_T B^2\over 6\pi m_e c}
\gamma_e^2 + {\gamma_e\over 3n}{dn\over dt'}\ .
\end{equation}
The first term on the right hand side of equation \ref{dgamma_dt}
represents the radiative losses while the second term represents
adiabatic cooling.  The radiative term includes only synchrotron
losses. A simple prescription of how to include the effects of
enhanced electron cooling, due to inverse compton scattering, on the
observed synchrotron emission, is given in \S \ref{dis}.

We use the BM spherical self-similar solution for an impulsive
explosion, where the external medium is cold and its density changes
as a power law of the distance from the center, $\rho_{\rm ext}(r)=Ar^{-k}$,
$k<4$ (extensions for $k>4$ are given in Best and Sari 2000, but were
not used in this paper). The derivations are made for a general value
of $k<4$, and are then used for $k=0$ and $k=2$, which are of special
physical interest. According to this solution, the proper energy
density, Lorentz factor and proper number density of the shocked fluid
are given by
\begin{eqnarray}
e&=&2\Gamma^2\rho_{\rm ext}c^2\chi^{-(17-4k)/3(4-k)}\label{BM_e}\ ,
\\
\gamma&=&2^{-1/2}\Gamma\,\chi^{-1/2}\label{BM_gamma}\ ,
\\
n&=&2^{3/2}\Gamma n_{\rm ext}\,\chi^{-(10-3k)/2(4-k)}\label{BM_n}\ ,
\end{eqnarray}
where $\Gamma$ is the Lorentz factor of the shock, and
\begin{equation}\label{chi}
\chi=\left[1+2(4-k)\Gamma^2\right]\left(1-{r\over ct}\right) \ .
\end{equation}
The $\chi$
coordinate of a fluid element is given by
\begin{equation}
\label{chi_to_t}
\chi=\left({R\over R_0}\right)^{4-k}=\left({t\over t_0}\right)^{4-k}
\end{equation}
where $R_0$ and $t_0$ are the shock radius and coordinate time,
respectively, when the fluid element crosses the shock. Since
$\Gamma^2\propto t^{k-3}$, we obtain that
\begin{equation}
\label{to_chi}
{\gamma\over\gamma_0}=\chi^{-{(7-2k)\over2(4-k)}} \quad,\quad {n\over
  n_0}=\chi^{-{(13-2k)\over 2(4-k)}}
\quad,\quad {e\over e_0}=\left({B\over B_0}\right)^2=\chi^{-{2(13-2k)\over 3(4-k)}} \ .
\end{equation}
Using equation \ref{to_chi} and the relation $dt'=dt/\gamma$, we can
write equation \ref{dgamma_dt} in terms of $\chi$:
\begin{equation}\label{dgamma_dchi} 
{d\gamma_e\over d\chi}=\, - \, {\sigma_T B_0^2 t_0 
\chi^{-(49-8k)/6(4-k)}\gamma_e^2\over 
6(4-k)\pi m_e c \gamma_0 }
\,  - \, {(13-2k)\over 6(4-k)}{\gamma_e\over\chi}\ .
\end{equation}
Solving equation \ref{dgamma_dchi} we obtain
\begin{equation}
\label{gamma_e_chi}
\gamma_{e}(\gamma_{e,0},\chi)={\gamma_{e,0}
  \over\chi^{(13-2k)/6(4-k)}+\gamma_{e,0} / \gamma_{\rm max}(\chi)} \ ,
\end{equation}
where $\gamma_{e,0}\equiv\gamma_e(\chi=1)$ is the initial Lorentz
factor of the electron, just behind the shock, and
$\gamma_{\rm max}(\chi)$ is the maximal Lorentz factor at $\chi>1$, which
corresponds to an electron with $\gamma_{e,0}\to \infty$, and is given
by
\begin{equation}\label{gamma_e_max}
\gamma_{\rm max}(\chi)={2(19-2k)\pi m_e c \gamma_0 \over 
\sigma_T B_0^2 t_0}\left({\chi^{(25-2k)/6(4-k)}\over \chi^{(19-2k)/3(4-k)}-1}\right)\ .
\end{equation}
The fraction of electrons with a Lorentz factor within the interval
$[\gamma_{e},\gamma_{e}+ d\gamma_{e}]$ is given by:
$N(\gamma_{e})d\gamma_{e}/n$, and remains constant as all these
quantities evolve with increasing $\chi$. The electron distribution is
therefore given by:
\begin{equation}\label{dist}
N(\gamma_{e},\chi)=
\end{equation}
$$
K_0 \chi^{(2k-13)(p+2)\over 6(4-k)}\gamma_{e}^{-p}
\left(1-{\gamma_{e}\over \gamma_{\rm max}(\chi)}
\right)^{p-2} \ \ 
{\rm for}\ \ \ \gamma_{\rm min}(\chi)\leq \gamma_{e}\leq \gamma_{\rm max}(\chi)
$$
where $\gamma_{\rm min}(\chi)=\gamma_e(\gamma_{{\rm min},0},\chi)$. 

We now have explicit expressions for both the hydrodynamical
quantities and the electron distribution, over all relevant
space-time, and can calculate the flux density near the various break
frequencies. For breaks that are in the optically thin regime
(b=2,3,9,11) one may use the equation
\begin{equation}\label{optically_thin}
F_{\nu}(t_{\rm obs})={2(4-k)R_l^3(1+z) \over d_{L}^2}\int^{1}_0dy\int^{y^{k-4}}_1 d\chi 
{\chi y^{2(5-k)}P'_{\nu'}(y,\chi,t_{\rm obs})\over \left[1+(7-2k)\chi y^{4-k}\right]^2} \ .
\end{equation}
which is a generalization of equation 13 of GPS99a, where $d_L$ and
$z$ are the luminosity distance and cosmological redshift of the
source, respectively, $P'_{\nu'}$ is the radiated power per unit
volume per unit frequency in the local rest frame of the fluid, and
should be taken at the coordinate time $t=t_z+r\mu/c$, where
$t_z\equiv t_{\rm obs}/(1+z)$,
\begin{equation}\label{RlGl}
R_l=\left[{(17-4k)(4-k)E t_z\over 4\pi A c}\right]^{1/(4-k)}\ ,
\end{equation}
$$
\gamma_l=\left[{(17-4k)E \over 
4^{5-k}(4-k)^{3-k}\pi A c^{5-k} t_z^{3-k}}\right]^{1/2(4-k)}\ ,
$$
$E$ is the energy of the blast wave, $y\equiv R/R_l$ (e.g. GPS99a),
\begin{equation}\label{mu}
\mu \equiv \cos(\theta) \cong 1-{1-\chi y^{4-k}\over 4(4-k)\gamma_l^2 y} \ ,
\end{equation}
and\footnote{Here, $\beta$ is the fluid velocity in units of the speed
  of light, rather than the spectral index.}
$\nu'=\nu\gamma(1-\beta\mu)$. The Spectral emissivity of a single
electron (in the fluid rest frame) is given by
\begin{equation}\label{Pnue}
P'_{\nu',e} = {\sqrt{3}q_e^3 B \sin\alpha\over m_e
  c^2}\, F\left({\nu'\over\nu'_{\rm syn}}\right)\quad,\quad
\nu'_{\rm syn} = {3 q_e B \gamma_e^2\sin\alpha \over 4\pi m_e c}\ ,
\end{equation}
where $q_e$ is the electric charge of the electron, $\alpha$ is the
pitch angle between the direction of the electron's velocity and the
magnetic field, in the local rest frame of the fluid, and $F$ is the
standard synchrotron function (e.g. Rybicki \& Lightman 1979).  In
order to obtain an expression for $P'_{\nu'}$ (which appears in
equation \ref{optically_thin}) we average $P'_{\nu',e}$ over
$\alpha$, assuming an isotropic distribution of electrons in the local
rest frame,
\begin{equation}\label{P_nu_e_ino}
P'_{\nu',e, {\rm iso}} = \int_0^{\pi/2}d\alpha \sin\alpha\, P'_{\nu',e}(\sin\alpha) \ ,
\end{equation}
and then integrate over the electron distribution,
\begin{equation}\label{Pnu}
P'_{\nu'} = \int_{\gamma_{\rm min}}^{\gamma_{\rm max}}d\gamma_e 
N(\gamma_e)P'_{\nu',e, {\rm iso}}(\gamma_e) \ .
\end{equation}

For the remaining spectral breaks (b=1,4,5,6,7,8,10), where the system
is not always optically thin, we follow the formalism of GPS99b.
Since the emission is isotropic in the local rest frame of the fluid,
the emission coefficient is simply $j'_{\nu'}=P'_{\nu'}/4\pi$, where
$P'_{\nu'}$ is given by equation \ref{Pnu}. The absorption coefficient
is given by
\begin{equation}\label{alpha_nu}
\alpha'_{\nu'} = {1\over 8\pi m_e \nu'^2}
\int_{\gamma_{\rm min}}^{\gamma_{\rm max}}d\gamma_e 
{N(\gamma_e)\over\gamma_e^2}{\partial\over\partial\gamma_e}
\left[\gamma_e^2 P'_{\nu',e, {\rm iso}}(\gamma_e)\right] \ .
\end{equation}
Since the flow is spherically symmetric, the afterglow image is
circular, with physical radius of
\begin{equation}\label{R_pmax}
R_{\perp,{\rm max}} = 
{(5-k)^{k-5\over 2(4-k)}\over \sqrt{2}}{R_l\over\gamma_l} =
\end{equation}
$$
\left[{2^{2-k}(17-4k)(4-k)^{5-k}E c^{3-k}t_z^{5-k}\over 
\pi(5-k)^{5-k}A}\right]^{1/2(4-k)}\ ,
$$
and for a given observer time, $t_{\rm obs}$, the specific intensity
(or brightness), $I_{\nu}$, depends only on the normalized radius from
the center of the image,
\begin{equation}\label{x}
x \equiv {R_{\perp}\over R_{\perp,{\rm max}}} = (4-k)^{-1/2}
(5-k)^{5-k\over 2(4-k)}\sqrt{y-\chi y^{5-k}}\ ,
\end{equation}
where $x=0$ at the center of the image and $x=1$ at the outer edge of
the image. As discussed in GPS99b, $I_{\nu}(x)$ may be obtained by
solving the radiative transfer equation,
\begin{equation}\label{rte}
{dI_{\nu}\over ds} = j_{\nu} - \alpha_{\nu}I_{\nu} \ ,
\end{equation}
where $s$ is the distance along the trajectory of a photon to the
observer, and the flux density is given by
\begin{equation}\label{F_nu_tau>1}
F_{\nu}(t_{\rm obs}) = 2\pi(1+z)\left[{R_{\perp,{\rm max}}(t_{\rm obs})\over d_L}
\right]^2\int_0^1 x\,dx\, I_{\nu}(x,t_{\rm obs})\ .
\end{equation}

We note that $I_{\nu}(x)$ provides the surface brightness profile of
the afterglow image, that is necessary for detailed calculations of
microlensing or scintillation. The surface brightness profiles that
were calculated according to this formalism, have already been used to
study the microlensing of GRB afterglows (Granot \& Loeb 2001; Gaudi,
Granot \& Loeb 2001), and are presented therein.

When a break frequency is sufficiently far from other
break frequencies, the spectrum near this break frequency assumes a
self similar form. These self similar forms of the spectrum near the
different break frequencies are presented in \S \ref{ANAP}.

\newcommand{\rb}[1]{\raisebox{1.5ex}[0pt]{#1}}
\begin{deluxetable}{llll}
  \tabcolsep0in \footnotesize \tablewidth{\hsize} \tablecaption{The
    Normalization of the Different Power Law Segments\label{PLS}}
  \tablehead{ \colhead{PLS} & \colhead{$\beta$} &
    \colhead{$F_{\nu}(k=0)$ in mJy} & \colhead{$F_{\nu}(k=2)$ in mJy}}
  \startdata A & $5/2$ &
  $1.18(4.59-p)10^{8}(1+z)^{9/4}\epsilon_B^{-1/4} n_0^{-1/2}
  E_{52}^{1/4}t_{\rm days}^{5/4}d_{L28}^{-2}\nu_{14}^{5/2}$ &
  $2.96(4.59-p)10^{7}(1+z)^{7/4}\epsilon_B^{-1/4}A_{*}^{-1}
  E_{52}^{3/4}t_{\rm days}^{7/4}d_{L28}^{-2}\nu_{14}^{5/2}$ \nl\hline
  B & $2$ & $4.20{(3p+2)\over(3p-1)}10^{9}(1+z)^{5/2} \bar\epsilon_{e}
  n_0^{-1/2} E_{52}^{1/2}t_{\rm days}^{1/2}d_{L28}^{-2}\nu_{14}^{2}$ &
  $1.33{(3p+2)\over (3p-1)}10^{9}(1+z)^{2}\bar\epsilon_{e}
  A_{*}^{-1}E_{52}t_{\rm days}d_{L28}^{-2}\nu_{14}^{2}$ \nl\hline C &
  $11/8$ & $8.01\cdot 10^{5}(1+z)^{27/16}\epsilon_B^{-1/4}n_0^{-5/16}
  E_{52}^{7/16}t_{\rm days}^{11/16}d_{L28}^{-2}\nu_{14}^{11/8}$ &
  $3.28\cdot 10^{5}(1+z)^{11/8}\epsilon_B^{-1/4}
  A_{*}^{-5/8}E_{52}^{3/4}t_{\rm days}d_{L28}^{-2}\nu_{14}^{11/8}$
  \nl\hline D & $1/3$ & $27.9{(p-1)\over(3p-1)}(1+z)^{5/6}
  \bar\epsilon_{e}^{-2/3}\epsilon_B^{1/3}n_0^{1/2}E_{52}^{5/6} t_{\rm
    days}^{1/2}d_{L28}^{-2}\nu_{14}^{1/3}$ &
  $211{(p-1)\over(3p-1)}(1+z)^{4/3}
  \bar\epsilon_{e}^{-2/3}\epsilon_B^{1/3}A_{*}E_{52}^{1/3}d_{L28}^{-2}\nu_{14}^{1/3}$
  \nl\hline E & $1/3$ & $73.0(1+z)^{7/6}\epsilon_B^{} n_0^{5/6}
  E_{52}^{7/6} t_{\rm days}^{1/6}d_{L28}^{-2}\nu_{14}^{1/3}$ &
  \nodata\tablenotemark{\dagger} \nl\hline F & $-1/2$ &
  $6.87(1+z)^{3/4}\epsilon_B^{-1/4}E_{52}^{3/4} t_{\rm
    days}^{-1/4}d_{L28}^{-2}\nu_{14}^{-1/2}$ &
  $6.68(1+z)^{3/4}\epsilon_B^{-1/4} E_{52}^{3/4}t_{\rm
    days}^{-1/4}d_{L28}^{-2}\nu_{14}^{-1/2}$ \nl\hline G & $(1-p)/2$ &
  $0.461(p-0.04)e^{2.53p}(1+z)^{3+p\over 4}
  \bar\epsilon_{e}^{p-1}\epsilon_B^{1+p\over
    4}n_0^{1/2}E_{52}^{3+p\over 4} t_{\rm days}^{3(1-p)\over
    4}d_{L28}^{-2}\nu_{14}^{1-p\over 2}$ &
  $3.82(p-0.18)e^{2.54p}(1+z)^{5+p\over 4}
  \bar\epsilon_{e}^{p-1}\epsilon_B^{1+p\over 4}A_{*}E_{52}^{1+p\over
    4} t_{\rm days}^{1-3p\over 4}d_{L28}^{-2}\nu_{14}^{1-p\over 2}$
  \nl\hline H & $-p/2$ & $0.855(p-0.98)e^{1.95p}(1+z)^{2+p\over 4}
  \bar\epsilon_{e}^{p-1}\epsilon_B^{p-2\over 4}E_{52}^{2+p\over 4}
  t_{\rm days}^{2-3p\over 4}d_{L28}^{-2}\nu_{14}^{-p/2}$ &
  $0.0381(7.11-p)e^{2.76p}(1+z)^{2+p\over 4}
  \bar\epsilon_{e}^{p-1}\epsilon_B^{p-2\over 4}E_{52}^{2+p\over 4}
  t_{\rm days}^{2-3p\over 4}d_{L28}^{-2}\nu_{14}^{-p/2}$ \enddata
  \tablecomments{The first two columns give the labels and the
    spectral slope, $\beta$, of the different PLSs (see Figure \ref{Fig1}),
    while the last two columns give the asymptotic flux density within
    each PLS, for $k=0$ and $k=2$. The reader is reminded that
    $\bar\epsilon_e=\epsilon_e(p-2)/(p-1)$ depends on $p$. The
    notation $Q_x$ stands for the quantity $Q$ in units of $10^x$
    times the (c.g.s) units of $Q$, while $t_{\rm days}$ is the
    observed time in days, and $A_{*}$ is $A$ in units of $5\cdot
    10^{11}\,{\rm gr/cm}$ (Chevalier \& Li 2000).}
    \tablenotetext{\dagger} {For PLS E, the emission becomes dominated
    by the contribution from small radii for $k>23/13$. This new
    regime is described in a separate work (Granot \& Sari, in
    preparation).}
\end{deluxetable}

\begin{deluxetable}{lcclllcc}
\tabcolsep0in
\footnotesize
\tablewidth{\hsize} \tablecaption{The Break Frequencies and
  Corresponding Flux Densities \label{T1}} \tablehead{ \colhead{$b$} &
  \colhead{$\beta_1$} & \colhead{$\beta_2$} & \colhead{$\nu_b$} &
  \colhead{$\nu_{b}(p)$ in Hz} & \colhead{$F_{\nu_{b},{\rm
        ext}}(p)$ in mJy} & \colhead{$s(p)$} & \colhead{MRD in \%}}
\startdata & & & & $1.24{(p-1)^{3/5}\over(3p+2)^{3/5}} 10^{9}
(1+z)^{-1}\bar\epsilon_{e}^{\; -1}\epsilon_B^{1/5} n_0^{3/5}
E_{52}^{1/5}$ & $0.647{(p-1)^{6/5}\over(3p-1)(3p+2)^{1/5}}
(1+z)^{1/2}\bar\epsilon_{e}^{\; -1}\epsilon_B^{2/5} n_0^{7/10}
E_{52}^{9/10}t_{\rm days}^{1/2}d_{L28}^{-2}$ & 1.64 & 6.68 \nl \rb{1} &
\rb{2} & \rb{\large${1\over 3}$} & \rb{$\nu_{sa}$} &
$8.31{(p-1)^{3/5}\over(3p+2)^{3/5}} 10^{9}
(1+z)^{-2/5}\bar\epsilon_{e}^{\; -1}\epsilon_B^{1/5} A_{*}^{6/5}
E_{52}^{-2/5}t_{\rm days}^{-3/5}$ & $9.19
{(p-1)^{6/5}\over(3p-1)(3p+2)^{1/5}}(1+z)^{6/5} \bar\epsilon_{e}^{\;
  -1}\epsilon_B^{2/5}A_{*}^{7/5} E_{52}^{1/5}
t_{\rm days}^{-1/5}d_{L28}^{-2}$ & 1.06 & 1.02 \nl \hline & & & &
$3.73(p-0.67) 10^{15} (1+z)^{1/2}E_{52}^{1/2}\bar\epsilon_{e}^2
\epsilon_B^{1/2}t_{\rm days}^{-3/2}$ & $9.93(p+0.14) (1+z)
\epsilon_B^{1/2} n_0^{1/2} E_{52} d_{L28}^{-2}$ & 1.84-0.40p & 5.9 \nl
\rb{2} & \rb{\large ${1\over 3}$} & \rb{\large ${1-p\over 2}$} &
\rb{$\nu_m$} & $4.02(p-0.69)
10^{15}(1+z)^{1/2}E_{52}^{1/2}\bar\epsilon_{e}^2 \epsilon_B^{1/2}
t_{\rm days}^{-3/2}$ & $76.9(p+0.12)(1+z)^{3/2} \epsilon_B^{1/2}
A_{*}E_{52}^{1/2}t_{\rm days}^{-1/2}d_{L28}^{-2}$ & 1.76-0.38p & 7.2 \nl
\hline & & & & $6.37(p-0.46) 10^{13}e^{-1.16p} (1+z)^{-1/2}
\epsilon_B^{-3/2}n_0^{-1} E_{52}^{-1/2}t_{\rm days}^{-1/2}$ & $4.68\,
e^{4.82(p-2.5)}10^3 (1+z)^{p+1\over 2}\bar\epsilon_{e}^{p-1}
\epsilon_B^{p-{1\over 2}} n_0^{p\over 2}E_{52}^{p+1\over
  2}t_{\rm days}^{1-p\over 2} d_{L28}^{-2}$ & 1.15-0.06p & 1.9 \nl \rb{3}
& \rb{\large ${1-p\over 2}$} & \rb{\large $-{p\over 2}$} &
\rb{$\nu_c$} & $4.40(3.45-p) 10^{10}e^{0.45p} (1+z)^{-3/2}
\epsilon_B^{-3/2}A_{*}^{-2} E_{52}^{1/2}t_{\rm days}^{1/2}$ & $8.02 \,
e^{7.02(p-2.5)} 10^{5} (1+z)^{p+{1\over 2}}\bar\epsilon_{e}^{p-1}
\epsilon_B^{p-{1\over 2}}A_{*}^{p}E_{52}^{1/2}t_{\rm days}^{{1\over 2}-p}
d_{L28}^{-2}$ & 0.80-0.03p & 4.4 \nl \hline & & & & $5.04(p-1.22)
10^{16} (1+z)^{1/2}\bar\epsilon_{e}^{2}\epsilon_{B}^{1/2}
E_{52}^{1/2}t_{\rm days}^{-3/2}$ & $3.72(p-1.79) 10^{15} (1+z)^{7/2}
\bar\epsilon_{e}^{5}\epsilon_{B}n_0^{-1/2}E_{52}^{3/2}t_{\rm days}^{-5/2}
d_{L28}^{-2}$ & $3.44p-1.41$\tablenotemark{\dagger} &
0.7\tablenotemark{\dagger} \nl \rb{4} & \rb{2} & \rb{\large ${5\over
    2}$} & \rb{$\nu_m$} & $8.08(p-1.22) 10^{16}
(1+z)^{1/2}\bar\epsilon_{e}^{2}\epsilon_{B}^{1/2}
E_{52}^{1/2}t_{\rm days}^{-3/2}$ & $3.04(p-1.79) 10^{15}
(1+z)^{3}\bar\epsilon_{e}^{5}\epsilon_{B}A_{*}^{-1}
E_{52}^{2}t_{\rm days}^{-2}d_{L28}^{-2}$ &
$3.63p-1.60$\tablenotemark{\dagger} & 1.8\tablenotemark{\dagger} \nl
\hline & & & & $3.59(4.03-p) 10^{9}e^{2.34p}
\left[{\bar\epsilon_{e}^{4(p-1)}\epsilon_{B}^{p+2}n_0^{4}E_{52}^{p+2}
    \over (1+z)^{6-p}t_{\rm days}^{3p+2}}\right]^{1/2(p+4)}$ &
$20.8(p-1.53) e^{2.56p} d_{L28}^{-2} \left[{
    (1+z)^{7p+3}\epsilon_{B}^{2p+3}E_{52}^{3p+7} \over
    \bar\epsilon_{e}^{10(1-p)} t_{\rm days}^{5(p-1)}}\right]^{1/2(p+4)}$ &
1.47-0.21p & 5.9 \nl \rb{5} & \rb{\large ${5\over 2}$} & \rb{\large
  ${1-p\over 2}$} & \rb{$\nu_{sa}$} &
$1.58(4.10-p)10^{10}e^{2.16p}\left[{
    \bar\epsilon_{e}^{4(p-1)}\epsilon_{B}^{p+2}A_{*}^{8} \over
    (1+z)^{2-p} E_{52}^{2-p}t_{\rm days}^{3(p+2)}} \right]^{1/2(p+4)}$ &
$158(p-1.48)e^{2.24p} d_{L28}^{-2} \left[{
    (1+z)^{6p+9}\epsilon_{B}^{2p+3}E_{52}^{4p+1} \over
    \bar\epsilon_{e}^{10(1-p)}
    A_{*}^{2(p-6)}t_{\rm days}^{4p+1}}\right]^{1/2(p+4)}$ & 1.25-0.18p &
7.2 \nl \hline & & & &
$3.23(p-1.76)10^{12}\left[{\bar\epsilon_{e}^{4(p-1)}
    \epsilon_{B}^{p-1}n_0^{2}E_{52}^{p+1} \over
    (1+z)^{7-p}t_{\rm days}^{3(p+1)}}\right]^{1/2(p+5)}$ &
$76.9(p-1.08)e^{2.06p}d_{L28}^{-2}\left[{
    (1+z)^{7p+5}\epsilon_{B}^{2p-5}E_{52}^{3p+5} \over
    \bar\epsilon_{e}^{10(1-p)}
    n_0^{p}t_{\rm days}^{5(p-1)}}\right]^{1/2(p+5)}$ & 0.94-0.14p & 12.4
\nl \rb{6} & \rb{\large ${5\over 2}$} & \rb{\large $-{p\over 2}$} &
\rb{$\nu_{sa}$} & $4.51(p-1.73)10^{12}\left[{\bar\epsilon_{e}^{4(p-1)}
    \epsilon_{B}^{p-1}A_{*}^{4}E_{52}^{p-1} \over (1+z)^{5-p}
    t_{\rm days}^{3p+5}}\right]^{1/2(p+5)}$ &
$78.6(p-1.12)e^{1.89p}d_{L28}^{-2}
\left[{(1+z)^{6p+5}\epsilon_{B}^{2p-5}E_{52}^{4p+5}\over
    \bar\epsilon_{e}^{10(1-p)}
    A_{*}^{2p}t_{\rm days}^{4p-5}}\right]^{1/2(p+5)}$ & 1.04-0.16p & 11.0
\nl \hline & & & &
$1.12{(3p-1)^{8/5}\over(3p+2)^{8/5}}10^{8}(1+z)^{-13/10}
\bar\epsilon_{e}^{-8/5}\epsilon_{B}^{-2/5}n_0^{3/10}E_{52}^{-1/10}t_{\rm days}^{3/10}$
& $5.27{(3p-1)^{11/5}\over(3p+2)^{11/5}}10^{-3}(1+z)^{-1/10}
\bar\epsilon_{e}^{-11/5}\epsilon_{B}^{-4/5}n_0^{1/10}
E_{52}^{3/10}t_{\rm days}^{11/10}d_{L28}^{-2}$ & 1.99-0.04p & 1.9 \nl
\rb{7} & \rb{2} & \rb{\large ${11\over 8}$} & \rb{$\nu_{ac}$} &
$1.68{(3p-1)^{8/5}\over(3p+2)^{8/5}}10^{8}(1+z)^{-1}
\bar\epsilon_{e}^{-8/5}\epsilon_{B}^{-2/5}A_{*}^{3/5}E_{52}^{-2/5}$ &
$3.76{(3p-1)^{11/5}\over(3p+2)^{11/5}}10^{-3}\bar\epsilon_{e}^{-11/5}
\epsilon_{B}^{-4/5}A_{*}^{1/5}E_{52}^{1/5}t_{\rm days}d_{L28}^{-2}$ &
1.97-0.04p & 1.9 \nl \hline & & & & $1.98 \cdot
10^{11}(1+z)^{-1/2}n_0^{1/6}E_{52}^{1/6}t_{\rm days}^{-1/2}$ &
$154(1+z)\epsilon_{B}^{-1/4}n_0^{-1/12}E_{52}^{2/3}d_{L28}^{-2}$ &
0.907 & 1.71 \nl \rb{8} & \rb{\large ${11\over 8}$} & \rb{\large
  $-{1\over 2}$} & \rb{$\nu_{sa}$} & $3.15 \cdot
10^{11}(1+z)^{-1/3}A_{*}^{1/3}t_{\rm days}^{-2/3}$ & $119(1+z)^{11/12}
\epsilon_{B}^{-1/4}A_{*}^{-1/6}E_{52}^{3/4}t_{\rm days}^{1/12}d_{L28}^{-2}$
& 0.893 & 2.29 \nl \hline & & & & $3.94(p-0.74)10^{15}(1+z)^{1/2}
\bar\epsilon_e^{\; 2} \epsilon_B^{1/2} E_{52}^{1/2}t_{\rm days}^{-3/2}$ &
$0.221(6.27-p) (1+z)^{1/2}\bar\epsilon_e^{\;-1}
\epsilon_B^{-1/2}E_{52}^{1/2}t_{\rm days}^{1/2}d_{L28}^{-2}$ & 3.34-0.82p
& 4.5 \nl \rb{9} & \rb{\large $-{1\over 2}$} & \rb{\large $-{p\over
    2}$}& \rb{$\nu_m$} & $3.52(p-0.31)10^{15} (1+z)^{1/2}
\bar\epsilon_e^{\; 2} \epsilon_B^{1/2} E_{52}^{1/2}t_{\rm days}^{-3/2}$ &
$0.165(7.14-p) (1+z)^{1/2}\bar\epsilon_e^{\;-1}
\epsilon_B^{-1/2}E_{52}^{1/2}t_{\rm days}^{1/2}d_{L28}^{-2}$ & 3.68-0.89p
& 4.2 \nl \hline 
& & & & $1.32 \cdot 10^{10}(1+z)^{-1/2}\epsilon_B^{6/5}n_0^{11/10}
E_{52}^{7/10}t_{\rm days}^{-1/2}$ & $3.72(1+z)\epsilon_B^{7/5}n_0^{6/5}
E_{52}^{7/5} d_{L28}^{-2}$ & 1.213 & 5.22 \nl \rb{10} & \rb{\large
  ${11\over 8}$} & \rb{\large ${1\over 3}$} & \rb{$\nu_{sa}$} & 
 \nodata\tablenotemark{\dagger\dagger} &
 \nodata\tablenotemark{\dagger\dagger} & 
\nodata\tablenotemark{\dagger\dagger} &
\nodata\tablenotemark{\dagger\dagger}\nl \hline
 &  & & & $5.86 \cdot 10^{12}(1+z)^{-1/2}\epsilon_B^{-3/2}n_0^{-1}
E_{52}^{-1/2}t_{\rm days}^{-1/2}$ & $28.4(1+z)\epsilon_B^{1/2}n_0^{1/2}
E_{52} d_{L28}^{-2}$ & 0.597 & 0.55 \nl 
\rb{11} & \rb{\large ${1\over 3}$} & \rb{\large $-{1\over 2}$} & \rb{$\nu_c$} &
\nodata\tablenotemark{\dagger\dagger} & 
\nodata\tablenotemark{\dagger\dagger} & 
\nodata\tablenotemark{\dagger\dagger} & 
\nodata\tablenotemark{\dagger\dagger} \nl 
\enddata 
\tablecomments{The first column numbers the breaks. The following 
  two columns are the asymptotic spectral slopes below ($\beta_1$) and
  above ($\beta_2$) the break. The fourth column gives the name of the
  break frequency.  The following two columns are $\nu_{b}(p)$ and
  $F_{\nu_b,{\rm ext}}(p)$. The last two columns are the parameter
  $s(p)$, which determines the shape of each break according to
  equation \ref{anapp} (except for $b=4$, where it applies to equation
  \ref{b4}), and the maximal relative difference (MRD) between this
  analytic formula and our exact numerical results. For each break
  frequency there are two lines, the first is for an ISM surrounding
  ($k=0$) and the second for a stellar wind environment ($k=2$). The
  reader is reminded that $\bar\epsilon_e=\epsilon_e(p-2)/(p-1)$
  depends on $p$.}  
\tablenotetext{\dagger} {For $b=4$, the values of
  $s(p)$ and the corresponding MRD refer to equation \ref{b4}, and not
  to equation \ref{anapp} as for the other breaks.}
\tablenotetext{\dagger\dagger} {The breaks $b=10,11$ involve PLS E,
  where the emission is dominated by the contribution from small radii
  for $k>23/13$. This new regime is described in a separate work
  (Granot \& Sari, in preparation).}
\end{deluxetable}

\newcommand{\rba}[1]{\raisebox{2.5ex}[0pt]{#1}}
\begin{deluxetable}{lcll}
  \tabcolsep0in \footnotesize \tablewidth{\hsize} \tablecaption{The
    Transition Times Between the Different Spectra\label{transitions}}
  \tablehead{ \colhead{${i \rightarrow j}$} & \colhead{possible definitions} &
 \colhead{$k$} & \colhead{transition time, $t_{i \rightarrow j}$, in days}}
  \startdata 
 & & 0 & $7.3\cdot 10^2-1.7\cdot 10^3\times(1+z)
\bar\epsilon_{e}^2\epsilon_{B}^2 n_0 E_{52}$ \nl 
\rb{$5\rightarrow 1$} & \rb{$\nu_2=\nu_3$, $\nu_9=\nu_{11}$, $\nu_7=\nu_{10}$} &
 2 & $2.0\cdot 10^2-7.0\cdot 10^2\times(1+z)\bar\epsilon_{e}\epsilon_{B}A_{*}$ \nl \hline 
 & & 0 & $6.1\cdot 10^4-1.2\cdot 10^6\times(1+z)\bar\epsilon_{e}^2
\epsilon_{B}^{1/5}n_0^{-2/5}E_{52}^{1/5}$ \nl 
\rb{$1\rightarrow 2$} & \rb{$\nu_1=\nu_2$, $\nu_4=\nu_5$} &
 2 & $1.2\cdot 10^7-3.9\cdot 10^9\times(1+z)
\bar\epsilon_{e}^{10/3}\epsilon_{B}^{1/3}A_{*}^{-4/3}E_{52}$ \nl \hline 
$4\rightarrow 5$ & $\nu_{10}=\nu_{11}$\tablenotemark{\dagger} & 2 &
$9.3(1+z)\epsilon_{B}^{9/7}A_{*}^{2}E_{52}^{-3/7}$ \nl\hline 
 & & 0 & $2.2\cdot 10^4-6.3\cdot10^5\times(1+z)
\bar\epsilon_{e}^2\epsilon_{B}^{1/2}n_0^{-1/6}E_{52}^{1/3}$ \nl 
\rb{$4\rightarrow 3$} & \rb{$\nu_4=\nu_6$, $\nu_7=\nu_8=\nu_9$} &
 2 & $1.5\cdot 10^5-1.1\cdot 10^7\times(1+z)\bar\epsilon_{e}^{12/5}
\epsilon_{B}^{3/5}A_{*}^{-2/5}E_{52}^{3/5}$ \nl \hline 
 & & 0 & $5.1\cdot10^8-1.2\cdot 10^9\times
(1+z)\bar\epsilon_{e}^2\epsilon_{B}^{2p+7\over p-1}
 n_2^{p+6\over p-1}E_{52}^{p+3\over p-1}$ \nl 
\rba{$3\rightarrow 2$} & \rba{$\nu_3=\nu_5$} &
 2 & $8.0-24\times(1+z)\bar\epsilon_{e}^{2(p-1)\over 2p+5}
\epsilon_{B}^{2p+7\over 2p+5}A_{*}^{2(p+6)\over 2p+5}E_{52}^{-{3\over 2p+5}}$ \nl 
\enddata
\tablecomments{The first column indicates the transition at hand,
  from spectrum $i$ to spectrum $j$.  The second column lists possible
  conditions that may be used to define the transition time.  The
  third column is $k$, which is either $0$ or $2$, for an ISM or
  stellar wind environment, respectively.  The last column is the
  transition time, $t_{i \rightarrow j}$. There are several different
  ways to define most of most transition times (see second column), resulting
  in numerical coefficients that differ by a factor of order unity.
  The $p$ dependence also varies the numerical coefficients by a
  factor of order unity. We specify the range of the numerical
  coefficients for $2.2<p<3$ and for the different definitions of each
  transition.}  
\tablenotetext{\dagger} {The expressions for $\nu_{10}$ and $\nu_{11}$
  for $k=2$, that we used in order to calculate $t_{4\rightarrow 5}$,
  are taken from Granot \& Sari (in preparation).}
\end{deluxetable}

\end{document}